\documentclass[10pt,conference]{IEEEtran}
\IEEEoverridecommandlockouts
\usepackage{cite}
\usepackage{amsmath,amssymb,amsfonts}
\usepackage{graphicx}
\usepackage{textcomp}
\usepackage{xcolor}

\usepackage{amsmath, amssymb, amsthm,algorithm}
\usepackage[latin1]{inputenc}
\usepackage{xcolor}
\usepackage{latexsym}
\usepackage{graphics,epsfig}
\usepackage{amsfonts}
\usepackage{booktabs}
\usepackage{color}
\usepackage{multirow}
\usepackage[justification=centering]{caption}
\usepackage{graphicx}
\usepackage{adjustbox}
\usepackage{kantlipsum}
\usepackage[caption=false,font=footnotesize]{subfig}
\usepackage{cases}
\usepackage{url}
\usepackage{tabu}
\usepackage{makecell}
\usepackage{soul}

\usepackage{array}
\usepackage{algpseudocode}
\usepackage[justification=centering]{caption}

\usepackage[caption=false,font=footnotesize]{subfig}
\usepackage[T1]{fontenc}
\usepackage{mwe}
\usepackage{subfig}
\usepackage{mathtools,lipsum,cuted}

\usepackage[english]{babel}
\usepackage{verbatim}

\usepackage[english]{babel}

\theoremstyle{plain}

\theoremstyle{remark}

\algrenewcommand{\algorithmiccomment}[1]{\hspace{-20px}$\rightarrow$ #1}

\def\BibTeX{{\rm B\kern-.05em{\sc i\kern-.025em b}\kern-.08em
    T\kern-.1667em\lower.7ex\hbox{E}\kern-.125emX}}
\begin{document}

\title{Maximizing Throughput with Routing Interference Avoidance in RIS-Assisted Relay Mesh Networks}
\author{\IEEEauthorblockN{Cao Vien Phung, Andre Drummond, and Admela Jukan}
\IEEEauthorblockA{Technische Universit\"at Braunschweig, Germany\\Email: \{c.phung, andre.drummond, a.jukan\}@tu-bs.de}}
%\author{\IEEEauthorblockN{Cao Vien Phung, Andre Drummond, and Admela Jukan}
%\IEEEauthorblockA{Technische Universit\"at Braunschweig, Germany \\
%Email: \{c.phung, andre.drummond, a.jukan\}@tu-bs.de}}
\maketitle

\begin{abstract}
In the modern landscape of wireless communications, multi-hop, high-bandwidth, indoor Terahertz (THz) wireless communications are gaining significant attention. These systems couple Reconfigurable Intelligent Surface (RIS) and relay devices within the emerging 6G network framework, offering promising solutions for creating cell-less, indoor, and on-demand mesh networks. RIS devices are especially attractive, constructed by an array of reflecting elements that can phase shifts, such that the reflecting signals can be focused, steered, and the power of the  signal enhanced towards the destination. This paper presents an in-depth, analytical examination of how path allocation impacts interference within such networks. We develop the first model which analyzes interference based on the geometric parameters of beams (conic, cylindrical) as they interact with  RIS, User Equipment (UE), and relay devices. We introduce a transmission scheduling heuristic designed to mitigate interference, alongside an efficient optimization method to maximize throughput. Our performance results elucidate the interference's effect on communication path quality and highlight effective path selection strategies with throughput maximization.

%\vspace{-0.3 cm}
 \end{abstract}
\begin{IEEEkeywords}
THz communications, Reconfigurable Intelligent Surface (RIS), mesh networks, routing, interference;
\end{IEEEkeywords}
%\vspace{-0.3 cm}

%%%%%%%%%%%%%%%%%%%%%%%%%%%%%%%%%%%%%%%%%%%%%%%%%%%%%%%%%%%%
%%%%%%%%%%%%%%%%%%%%%%%%%%%%%%%%%%%%%%%%%%%%%%%%%%%%%%%%%%%%
\section{Introduction} \label{intro}
\par Today, 6G-wireless networks are emerging rapidly due to the advances  in Terahertz (THz) communications and their extended reach of transmission. In THz frequency ranges, Reconfigurable Intelligent Surfaces (RISs) carry promise to mitigate issues of blockage due to Line of Sight (LoS) requirement, which is a common phenomena in indoor settings where people and objects move \cite{9565222, 9686640}. While 6G mesh THz networks are still in their infancy, applications such as private campus networks, or smart factories are driving their rapid evolution. In such scenarios, routing the wireless connections over multiple relays and RISs in a cell-less scenario between users and base stations is an open challenge.

\par
\begin{figure}[ht]
%\vspace{0.1 cm}
\centering
{\includegraphics[ width=7.9cm, height=4.5cm]{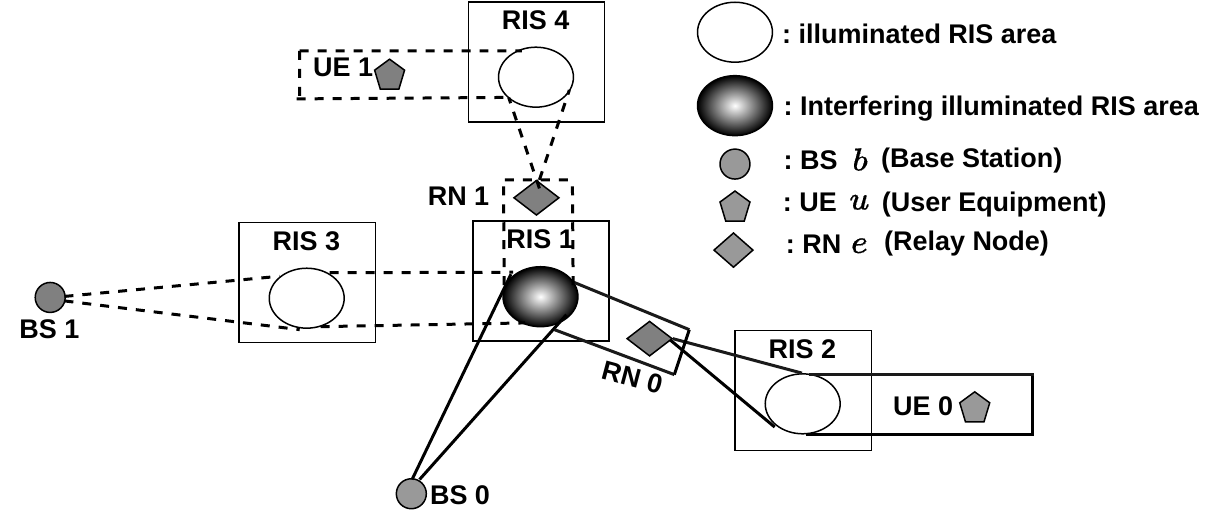}}
\caption{An example of interference on a RIS.} \label{example}
%\vspace{-0.68 cm}
\end{figure}

  In THz relay mesh networks, two transmissions interfere when their illuminated areas intersect. Consider an example shown in Fig. \ref{example}, with Base Stations (BSs), RISs, User Equipments (UEs) and Relay Nodes (RNs). Let us consider two transmissions: BS $1$ $\rightarrow$ RIS $3$ $\rightarrow$ RIS $1$ $\rightarrow$ RN $1$ $\rightarrow$ RIS $4$ $\rightarrow$ UE $1$ and BS $0$ $\rightarrow$ RIS $1$ $\rightarrow$ RN $0$ $\rightarrow$ RIS $2$ $\rightarrow$ UE $0$, that interfere on RIS $1$ (similar examples of interference can also be shown for relay nodes or UEs). Here, we assume that base stations use conical-, whereas RIS uses cylindrical beam shape.  As it can be seen, conical and cylindrical beam shapes interfere on a circular geometric coverage. The size of that coverage determines the impact of interference on signal quality, and path selection.

\par In this paper, we first analyze interference for beam steering under conical and cylindrical shapes, which can occur on any intermediate and receiving nodes, such as RISs, relay nodes, UEs. After providing a interference coverage analysis, we focus on routing and path selection methods for interference avoidance. To this end, we propose a novel path selection and optimization method, which we evaluate with simulations. In summary, our novel contributions are two-fold:
\begin {itemize}
\item {Analytical model of transmission beam shapes (conical or cylindrical), the resulting interference model in the network along with the derivation of RIS illuminated area as affected by beams from various  network nodes (BSs, other RISs, and relay nodes).}
%\item {Derivation and analysis of interference in THz relay mesh networks under various transmission parameters assumptions and the related throughput analysis}.
\item {An efficient scheduling transmission algorithms for the network system  to optimize throughput with various path computation methods, whereby paths can be optimized too, while the number of relay nodes allocated minimized.}
\end {itemize}

The remainder of the paper is organized as follows. Section \ref{relwor} discusses related work. Section \ref{analytical} describes the analytical framework of interference coverage. Section \ref{opti} presents path selection and optimization techniques. In section \ref{perev}, we show the performance evaluation. Section \ref{conl} concludes the paper.

%%%%%%%%%%%%%%%%%%%%% Related work
\section{Related work}\label{relwor}
\par Studies in \cite{9840504,9179528} deploy RIS-assisted relay millimeter wave communications, whereby relay nodes are used to support passive RISs to reducing the far-field path-loss. Our paper uses this relay mesh reference scenario here applied to THz relay mesh networks. In such networks, interference can be studied as it applies RISs, UEs and relay nodes, which previous work has addressed as follows. 

\par Paper \cite{10099527} analyses interference aware routing in millimeter wave mesh networks, however without consideration of RIS- and relay-assisted mesh networks. Papers \cite{10.1093/comjnl/bxaa083,7820226,9456082} model interference and coverage for indoor THz or millimeter wave communication beamforming antennas. We share the assumption from these studies that any receiver can suffer interference from any users. We furthermore extend this body of work to include the interference from intermediate nodes, e.g., relay nodes and RISs, and we also consider the volume of beams that are actually causing interference. 

\par As analyzed in \cite{9145080}, a THz mesh network with uncoordinated directional transmissions can cause a significant interference for receiving users. Paper \cite{9049790} proposes methods for transmission scheduling among directional links in 6G THz mesh networks to fully exploit concurrent transmissions, thus avoiding cross-link interference, which is also our goal. \cite{9840504} analyzes  the interference on relay nodes in RIS- and relay-assisted millimeter wave communications, wheras is focus on THz communication model, which is different.  Study in \cite{9681803} focuses on interference on an RIS-enabled multiple-access environment modeled as $K$-user (multi-user) interference channel to select RIS reflection coefficients on one RIS, so that the system could achieve interference nulling, when the number of RIS reflecting elements is slightly larger than $2K(K-1)$. However, this approach is only deployed in a two-hop scenario with only one RIS. For RIS-assisted multi-hop mesh networks this is still an open issue because multi-hop transmissions include one or more RISs. Hence, selecting a RIS with reflection coefficients to perform nulling interference for each transmission to maximize the number of paths with interference  nulling is not straightforward. %The selection of RIS reflection coefficients is here out of scope.

\par In \cite{9515976}, a schematic representation of RIS-aided beam steering is presented, where the incident beam is reflected by RIS  as a beam along the direction with transmission angles defined. We use this characteristics of RISs to analyze the beam shapes, which in THz networks are cylinders reflected from RISs. Paper \cite{7820226} analyzes conical beams generated from transmitters, which we use to analyze beam shapes generated from BSs/RNs. Paper \cite{9386246} found that illuminated areas captured by RISs from the conical beams generated by transmitters are circular ones, which we use in our model here. Also, based on the method of scheduling transmissions proposed in \cite{viennew}, we make significant changes to its usage, since THz communications use directional- and not omni antennas. The distinction is important for conflict free transmission scheduling over paths, because THz communication have high path loss, so any relay node/RIS/UE within THz interference range cannot be intermediately counted as a conflict. In our algorithm, we consider SNR at UE/relay nodes to decide whether the transmission is conflicted.

\par In terms of throughput analysis, our focus is on  multi-hop RIS-assisted THz networks which we base on \cite{9410457},  which improves the coverage range in THz band. We use this approach to mitigate the effect of longer distances or blockage. We furthermore borrow from the transmission approach proposed in \cite{5438834}, where the throughput is defined as a maximum multiplier $\lambda$ such that each demand with its traffic value is multiplied by $\lambda$, then it can be feasibly routed in the network. Paper \cite{5438834} shows how to maximize the throughput, which is equivalent to maximizing $\lambda$, by finding suitable paths. Central to our analysis is the adoption of the value $\lambda$ to solving the problem of maximizing throughput with routing avoidance. 

%%%%%%%%%%%%%%%%%%%%%%% Analysis
\section{Analytical model} \label{analytical}

In this section, we first describe the reference scenario. After that we provide the baseline SNR analysis for THz transmission. We finish with interference analysis. 

%%%%%%%%%%% reference scenario
\subsection {Reference scenario} \label{refsce} 

\par
\begin{figure*}[ht]
 %\vspace{0.1 cm}
\centering
  {\includegraphics[ width=16cm, height=7.49cm]{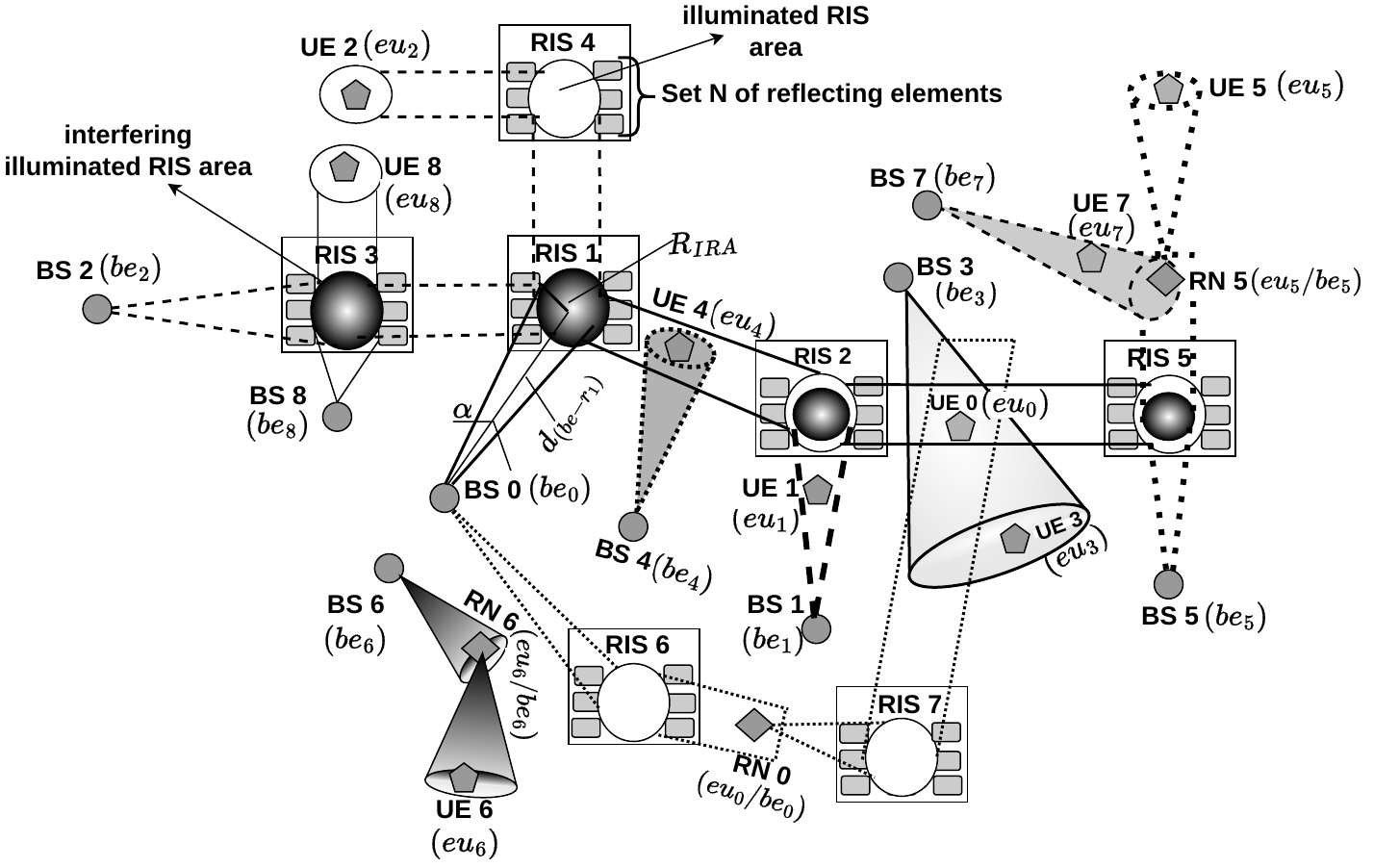}}
  \caption{ Reference scenario for the analysis, where BS $b$ (Base Station), UE $u$ ((User Equipment)), RN $e$ (Relay Node), $\alpha$ (directivity angle), $be$ (transmitting node BS or RN), $eu$ (receiving node RN or UE), $R_{IRA}$ (radius of illuminated area), $d_{be-r_1}$ (distance from $be$ to the first RIS $r_1$ of any transmission), RIS (Reconfigurable Intelligent Surface). } \label{genscenario}
  %\vspace{-0.68 cm}
  \end{figure*}

 Fig. \ref{genscenario} illustrates a THz network that consists of $B$ base stations (referred to as source nodes), $R$ passive RISs (here corresponding to intermediate nodes), $E$ relay nodes (RNs) (here corresponding to intermediate nodes), and  $U$ user equipment devices (UE,  as destination nodes). Each RIS is designed with $N$ configurable reflecting elements. We assume that BSs always act as transmitters, while UEs as receivers.  RISs and RNs on the other hand can both transmit and receive data. RISs are passive elements, while the RNs act as repeaters in multi-hop network and can be used to improve SNR \cite{9840504}. Based on the properties of THz communications and RIS devices, we assume the following beam shapes:  a) conical from BSs/RNs \cite{7820226}, b)  cylindrical  from RISs \cite{9515976}, and c) circular, the areas  on a RIS illuminated by the conical beams from BSs/RNs \cite{9386246}. 

\par Consider an example in Fig. \ref{genscenario}, where BS $0$ sends its data to UE $0$. Out of two possible paths, BS $0$  chooses path BS $0$ $\rightarrow$ RIS $6$ $\rightarrow$ RN $0$ $\rightarrow$ RIS $7$ $\rightarrow$ UE $0$ over the path BS $0$ $\rightarrow$ RIS $1$ $\rightarrow$ RIS $2$ $\rightarrow$ UE $0$. This is due to the latter path interfering more with other active paths. For instance, path BS $2$ $\rightarrow$ RIS $3$ $\rightarrow$ RIS $1$ $\rightarrow$ RIS $4$ $\rightarrow$ UE $2$ can interfere on RIS $1$ with the path coming from BS $0$. Similarly, path BS $1$ $\rightarrow$ UE $1$ interferes on RIS $2$ with the path coming from RIS $1$. Other paths also interfere in their corresponding beam shape volume, such as cylindrical and conical.

Relay nodes are used on multi-hop paths when the  Signal-to-Noise Ratio (SNR) at UEs is $SNR \leq T$ (i.e., threshold value accepted in THz transmission). We assume that any RNs are chosen such that the receiving $SNR>T$. For instance, assuming that path BS $0$ $\rightarrow$ RIS $6$ $\rightarrow$ RN $0$ $\rightarrow$ RIS $7$ $\rightarrow$ UE $0$ has not only $SNR>T$ at UE $0$, but also $SNR>T$ at RN $0$.  On the other hand, RNs might also cause blocking if two paths attempt to allocate it. For example, if path BS $0$ $\rightarrow$ RIS $6$ $\rightarrow$ RN $0$ $\rightarrow$ RIS $7$ $\rightarrow$ UE $0$ uses RN $6$,  path BS $6$ $\rightarrow$ RN $6$ $\rightarrow$ UE $6$ can be blocked. The placement and sharing of RNs are an important consideration that deserves a separate study. In our approach, we propose an algorithm to allocating relay nodes RNs such that the number of relay nodes used for each path can be minimized, while satisfying the end-to-end SNR path constraint. 

\subsection{End-to-end Signal-to-Noise Ratio (SNR)} \label{SNRend}
%\hl{this section should be merged with scheduling section since it is ONLY used there and nowehere else}
\par Let us define a THz data transmission channel between one BS and one UE/relay node, or between one relay node and UE/relay node. Based on \cite{9840504}, we extend the SNR analysis to include  characteristics of RISs and THz communications between one transmitting node $be \in \{B \cup E \}$ (i.e., BS $b$ or RN relay node $e$) and one receiving node $eu \in \{E \cup U \}$ (i.e., relay node $e$ or UE $u$) via a set $I_{(be,eu)}$ of RISs $r \in R$, i.e.,
\begin{equation} \label{calZ}
SNR_{(be,eu)} = \frac{P_{eu} \cdot G_{be} \cdot G_{eu}}{k\cdot T\cdot W}
\end{equation}

Eq. \eqref{calZ} uses noise power $P_{noise}=k\cdot T \cdot W$, where $k$ is the Boltzmann constant, $T$ denotes the absolute temperature in Kelvin, and $W$ is the bandwidth. $G_{be}$ denotes the transmitting antenna gain of node $be$, while $G_{eu}$ is the receiving antenna gain of node $eu$. The transmitting antenna gain $G_{be}$ with beamwidth $\alpha$, as illustrated with a directivity angle in Fig. \ref{genscenario} from BS $0$ towards RIS $1$, is given by \cite{7820226}:
\begin{equation}\label{Gbe}
G_{be}=\frac{2}{1-cos(\frac{\alpha}{2})},
\end{equation}
and we assume that $G_{be}=G_{eu}$. For the optimal phase to maximize the transmission speed of each RIS $r$ \cite{8796365}, the phase shift of each reflecting element of RIS with respect to the related network node is similar to the phase shift of the channel between the reflecting element of that RIS and that related network node. So, for case of $|I_{(be,eu)}|=1$, the signal (without considering $G_{be}$ and $G_{eu}$) obtained at node $eu$ is given by:
\begin{equation} \label{eqex1}
\begin{split}
P_{eu}= P_{be}  | H_{(be,r_1)} \cdot |N^\prime| \cdot H_{(r_1,eu)} |^2,
\end{split}
\end{equation}
where $N^\prime$ is the set of illuminated reflecting elements (a subset of $N$). For $|I_{(be,eu)}|>1$, the signal (without considering $G_{be}$ and $G_{eu}$) obtained at node $eu$ is given by:
\begin{equation} \label{eqex2}
\begin{split}
& P_{eu} = P_{be}  \left | (H_{(r_{|I_{(be,eu)}|},eu)} \cdot |N^\prime|)  \times \right. \\ & \left. \left ( \prod_{i=2}^{|I_{(be,eu)}|-1} H_{ (r_i,r_{i+1}) } \cdot |N^\prime| \right) \times   (H_{(be,r_1)} \cdot H_{(r_1,r_2)} \cdot |N^\prime|)\right |^2
\end{split}
\end{equation}
where $P_{be}$ is transmitting power of node $be$;  channel transfer function $H_{(be,r_1)}$ between the transmitting node $be$ and the $1^{st}$ RIS receiving node $r_1$ of any transmission, $H_{(r_i,r_{i+1})}$ between the $i^{th}$ RIS transmitting node $r_i$ and the $(i+1)^{th}$ RIS receiving node $r_{i+1}$, $H_{(r_{|I_{(be,eu)}|},eu)}$ between the $|I_{(be,eu)}|^{th}$ RIS transmitting node $r_{|I_{(be,eu)}|}$ and the receiving node $eu$, are expressed by a common function:
\begin{equation} \label{commonH}
\begin{split}
H=\left( \frac{c}{4\cdot \pi \cdot f \cdot d} \right)e^{(-\frac{1}{2}k(f) \cdot d)},
\end{split}
\end{equation}
where $c$ is the speed of light, $f$ denotes the operating THz frequency, $d$ represents the transmission distance between two network nodes, and $k(f)$ stands for overall molecular absorption coefficients of the medium at THz band. Finally,  the SNR of one transmission between a pair of $be$ and $eu $ without RIS ($|I_{(be,eu)}|=0$) can be expressed as:
\begin{equation} \label{SNRwoRIS}
\begin{split}
SNR_{(be,eu)} = \frac{P_{be} |H_{(be,eu)}|^2 \cdot G_{be} \cdot G_{eu}}{k\cdot T\cdot W},
\end{split}
\end{equation}
where $H_{(be,ue)}$ can be referred to Eq. \eqref{commonH}, and
\begin{equation} \label{peuworis}
P_{eu}=P_{be}|H_{(be,eu)}|^2.
\end{equation} 

\subsection{Interference Analysis} \label{beamshape}
Let us define the number of illuminated RIS reflecting elements RIS as $N^\prime $, a subset of $N$. If BS $b$ or RN relay node $e$ emit their signals with antenna directivity angle $\alpha$, the corresponding beam shape is conical, e.g., node BS $0$ and RN 0 towards RIS $1$ and RIS $7$, respectively, in Fig. \ref{genscenario}; the illuminated RIS area is circular, as discussed in \cite{9386246}. Since to illuminated RIS area is circular, when reflecting signals from other UE/RN/RIS the resulting beam shape is approximated as cylindrical. For instance, RIS $1$ reflects its signal to RIS $2$ in a cylinder beam shape in Fig. \ref{genscenario}. Furthermore, we assume that all actual illuminated areas on all RISs of the same transmission have the same radius $R_{IRA}$. The radius $\mathbb{R}_{fp}$ of the footprint of conical shape can be approximately calculated by using trigonometric ratio in right-angled triangles:
\begin{equation} \label{radius}
 \mathbb{R}_{fp}=tan\left(\frac{\alpha}{2}\right)\cdot d_{(be-r_1)},
\end{equation}
where $d_{(be-r_1)}$ is the transmission distance between $be$ and the $1^{st}$ RIS $r_1$ of any transmission, and $be$ denotes BS $b$ and RN relay node $e$, i.e., $be \in \{B \cup E \}$. The area $S_{fp}$ of the footprint of conical shape can be approximately given by:
\begin{equation} \label{Sira}
S_{fp} \approx \pi \cdot \mathbb{R}_{fp}^2.
\end{equation}
Let us assume that all RISs have the same size of plane area  $S_{RIS}$ and the same radius $R_{RIS}$, whereby each plane area of a RIS contains $N$  reflecting elements, which in practice amounts to hundreds or even thousands. If $S_{fp} \leq S_{RIS}$, the entire power from one transmitting node $be$ can be reflected by the RIS towards the other RISs or network nodes $eu$ because the entire footprint of its beam can be captured by the RIS. Otherwise, a part of incident power from one transmitting node $be$ can be lost because it resides beyond the RIS surface, i.e., the footprint size of the beam is larger than the size of RIS plane. Therefore, the actual illuminated RIS area is:
\begin{equation} \label{Sfinal}
S=min(S_{fp}, S_{RIS}).
\end{equation}

The radius of actual illuminated RIS area is given as:
\begin{equation}\label{Ractualill}
R_{IRA}=min(\mathbb{R}_{fp},R_{RIS})
\end{equation}

The number of illuminated reflecting elements of each RIS can be expressed by:
\begin{equation} \label{Ncomma}
|N^\prime|=\frac{S}{dx \cdot dy} \leq N,
\end{equation}
where $dx$ and $dy$ denotes the $x$ and $y$ dimensions of each reflecting element. We model the conical and cylindrical beam coverage of any transmissions over RISs as:
\begin{subnumcases}{V=}
  \frac{1}{3} \cdot \pi \cdot \mathbb{R}_{fp}^2 \cdot d_i  &, if $i=1$, \label{cone1}\\
    \pi \cdot R_{IRA}^2 \cdot d_{i}  &, if $1 < i < h$, \label{cylinder1}\\
 \pi \cdot R_{IRA}^2 \cdot d_{h}  &, if $i=h$, \label{cylinder2} 
\end{subnumcases}
where Eq. \eqref{cone1} denotes the conical volume of transmission beam from the first hop, i.e., node $be$ to the first RIS $r_1$ of transmission,  $d_i$ is the transmission distance of that hop, and the radius $\mathbb{R}_{fp}$ of the footprint of conical shape is measured by Eq. \eqref{radius}. Eq. \eqref{cylinder1} is the cylindrical volume for the hop $i$ between two RISs of the transmission, where $h$ is the number of hops of transmission, and the radius $R_{IRA}$ of actual illuminated RIS area is applied from Eq. \eqref{Ractualill}. Eq. \eqref{cylinder2} is the cylindrical volume for the last hop, i.e.,
\begin{equation}\label{lasthop}
d_{h} = d_{th} -  \sum_{j=1}^{h-1} d_j,
\end{equation}
where $d_{th}$ is the threshold distance, whereby there is no interference at nodes with transmission distances farther than the threshold distance. With threshold  $T$ which node $eu$ detects as the signal coming from node $be$, and replacing $SNR_{(be,eu)}$ by $T$ from Eq. \eqref{calZ} and Eq. \eqref{SNRwoRIS}, we can calculate $d_{th}$. 

\par Finally, a conical beam coverage of a path connecting directly between transmitting node $be$ and receiving node $eu$ (e.g., BS $6$ $\rightarrow$ RN $6$ or RN $6$ $\rightarrow$ UE $6$ in Fig. \ref{genscenario}) is given by Eq. \eqref{cone1} (equation for calculating conical beam coverage), where $d_i = d_{th}$. 

Based on the characteristics of the narrowly directional beamforming of THz communications and RISs, as illustrated in Fig. \ref{genscenario}, we identify two typical cases of the beam that can cause interference, on an UE, RN or on a RIS:
\begin{itemize}
\item \textbf{Conical}: Interference by beams with conical shape. 

Two illustrative cases of this interference are shown in Fig. \ref{genscenario}, one on nodes, and the other on RISs. For instance,  BS $3 \rightarrow$ UE $3$ causes interference for path BS $0 \rightarrow$ RIS $1 \rightarrow$ RIS $2 \rightarrow$ UE $0$. Similarly, on RIS $2$ the conical beam BS $1 \rightarrow$ UE $1$ interferes with the path BS $0 \rightarrow$ RIS $1 \rightarrow$ RIS $2 \rightarrow$ UE $0$. 

\item \textbf{Cylindrical}:  Interference of the beams with cylindrical shape. For instance, in Fig. \ref{genscenario}, BS $0 \rightarrow$ RIS $1 \rightarrow$ RIS $2 \rightarrow$ UE $0$ interferes on node UE $4$ with BS $4 \rightarrow$ UE $4$ . Similarly, on RIS $5$ path BS $5 \rightarrow$ RIS $5 \rightarrow$ RN $5 \rightarrow$ UE $5$ experiences interference from BS $0 \rightarrow$ RIS $1 \rightarrow$ RIS $2 \rightarrow$ UE $0$
\end{itemize}

For a given network topology modeled as a graph $ \mathbb{G}=( \mathbb{N},\mathbb{E})$, a path (defined in \ref{SNRend}) with traffic $y_{(be,eu)}$ sent from node $be$ to node $eu$ is denoted as $(be,\{I_{(be,eu)}\cup eu\})$, whereby $I_{(be,eu)}$ is the set of RISs in the transmission between node $be$ and node $eu$, where $|I_{(be,eu)}| \geq 0$. When $|I_{(be,eu)}| = 0$, the path does not contain RIS. Let us now analyze various cases of path interference for a pair $(be,\{I_{(be,eu)}\cup eu\})$ and $(be^\prime,\{I^\prime_{(be,eu)}\cup eu^\prime\})$ and the resulting interference,  when a node $i \in \{ I_{(be,eu)}\cup eu \}$ is within transmission coverage of $(be^\prime,\{I^\prime_{(be,eu)}\cup eu^\prime\})$, and vice-versa.  In cases without RISs, we use Eq. \eqref{cone1}, Eq. \eqref{cylinder1}, and Eq. \eqref{cylinder2} for the model of conical and cylindrical coverage in subsection \ref{beamshape}, possibly causing interference from any transmissions with RISs $(be,\{I_{(be,eu)}\cup eu\})$, where $|I_{(be,eu)}|>0$. The conical coverage of a transmission $(be,eu)$ without RIS is given by Eq. \eqref{cone1}, where $d_i = d_{th}$. 

Let us assume that the conical and cylindrical beam shapes are always from the node $be$ and from the RIS, respectively. We use Eq. \eqref{cone1}, Eq. \eqref{cylinder1}, and Eq. \eqref{cylinder2} to model of conical and cylindrical coverage in Section \ref{beamshape}, to obtain interference they cause from a path that includes RISs $(be,\{I_{(be,eu)}\cup eu\})$, where $|I_{(be,eu)}|>0$. As previously mentioned, the conical coverage of a transmission $(be,eu)$ that does not include RISs is given by Eq. \eqref{cone1}, where $d_i = d_{th}$. 

For conical shape with RISs, the threshold distance $d_{th}$ of transmission $(be,eu)$ (e.g. Fig. \ref{genscenario}, BS $1 \rightarrow$ UE $1$) causing interference is larger than or equal to the distance between node $be$ (Fig. \ref{genscenario}, BS $1$) of that transmission and the RIS of the interfered transmission (Fig. \ref{genscenario},  RIS $2$). Assuming that $N^\prime$ is the set of illuminated reflecting elements on RIS from the interfered transmission and $N^{\prime\prime}$ is the set of illuminated reflecting elements on that RIS from the transmission causing interference, then $|N^\prime|$ and $|N^{\prime\prime}|$ can be derived from Eq. \eqref{radius}, Eq. \eqref{Sira}, Eq. \eqref{Sfinal}, and Eq. \eqref{Ncomma}. The set of illuminated reflecting elements actually affected by the interference is here  given by:
\begin{equation}\label{interectioninter}
N_{ \imath } = N^\prime \cap N^{\prime\prime}.
\end{equation}

For cylindrical case with RISs, on the other hand, we can similarly analyze the set of illuminated reflecting elements as they interfere, whereby the distance $d_{h}$ (Eq. \eqref{lasthop}) of the last hop of the transmission causing interference is larger than or equal to the distance between the last RIS of the beam causing interference (e.g. Fig. \ref{genscenario}, RIS $2$) and the RIS of the path interfered (Fig. \ref{genscenario}, RIS $5$). 

\par We observe that for cases with RISs, based on Eq. \eqref{interectioninter}, the  path affected by interference experiences a partial coverage by the interfering beam. As analyzed in Section \ref{SNRend}, the radius $R_{IRA}$ of illuminated areas of RISs will not change during transmission over different RISs. Therefore, the illuminated areas actually interfered also do not change in size during transmission over different RISs.

\par If path from $(be,\{R_{(be,eu)}\cup eu\})$ is within the transmission coverage of $(be^\prime,\{R^\prime_{(be,eu)}\cup eu^\prime\})$,  then the Signal-to-Noise-plus-Interference Ratio (SNIR) of $(be,\{R_{(be,eu)}\cup eu\})$ is:
\begin{equation} \label{SNIRvi}
SNIR_{(be,eu)} = \frac{P_{eu} \cdot G_{be} \cdot G_{eu}}{k \cdot T \cdot W + \mathbb{I}_{(be^\prime,eu)}}, 
\end{equation} where $\mathbb{I}_{(be^\prime,eu)}$ denotes the interference of node $be^\prime$ of $(be^\prime,\{R^\prime_{(be,eu)}\cup eu^\prime\})$ to node $be$ of $(be,\{R_{(be,eu)}\cup eu\})$, and is given by:
\begin{equation}\label{interequ}
\mathbb{I}_{(be^\prime,eu)} = \delta_{eu} \cdot G_{be^\prime} \cdot G_{eu},
\end{equation} where $\delta_{eu}$ is the interference (without antenna gains $G_{be^\prime}$ and $G_{eu}$) on node $be$.  For cases in which $(be^\prime,\{R^\prime_{(be,eu)}\cup eu^\prime\})$ causes the interference on RISs, $(be,\{R_{(be,eu)}\cup eu\})$ is interfered partly or entirely by an undesired beam $(be^\prime,\{R^\prime_{(be,eu)}\cup eu^\prime\})$, so $\delta_{eu}$ is similarly calculated as Eq. \eqref{eqex1}, Eq. \eqref{eqex2}, but $N^\prime$ is replaced by $N_{ \imath } $ of Eq. \eqref{interectioninter}. We note that the path causing interference includes the related nodes  $(be,\{R_{(be,eu)}\cup eu\})$ and  $(be^\prime,\{R^\prime_{(be,eu)}\cup eu^\prime\})$. For instance, in Fig. \ref{genscenario}, one path causing interference is also: BS $2$ $\rightarrow$ RIS $3$ $\rightarrow$ RIS $1$ $\rightarrow$ RIS $2$ $\rightarrow$ UE $0$. For cases in which $(be^\prime,\{R^\prime_{(be,eu)}\cup eu^\prime\})$ causes directly the interference on the node $eu$ (relay node $e$ or UE node $u$) of $(be,\{R_{(be,eu)}\cup eu\})$, $(be,\{R_{(be,eu)}\cup eu\})$ is interfered by the whole undesired beam, so $\delta_{eu}$ is similarly calculated as Eq. \eqref{eqex1}, Eq. \eqref{eqex2}, and Eq. \eqref{peuworis}.

\par In summary, our analytical model of transmission beam shapes (conical or cylindrical) provides the basis for assessing the interference model in the network along with the derivation of RIS illuminated area as affected by beams from various  network nodes (BSs, other RISs, and relay nodes). In terms of interference's effect on RN and UE we postulate that when RN relay node $e$ or UE node $u$ of any transmission $(be,\{R_{(be,eu)}\cup eu\})$ is aftected by the interference from other paths $(be^\prime,\{R^\prime_{(be,eu)}\cup eu^\prime\})$, we assume that they are  captured the entire interfering beam shape. Therefore, this interference can simply be derived by Eq. \eqref{interequ} and Eq. \eqref{eqex1}, or Eq. \ref{eqex2}.

%%%%%%%%%%%%%%%%%%%%%%%%%%% path selection section

\section{Path Selection and Optimizations}\label{opti}

\par The path selection and optimization happens in three main steps. In the first step, all candidate paths are selected that satisify $SNR > T$. In addition, other paths characterized by low quality,  i.e., $SNR \leq T$ can also be included, but need to include relay nodes (RN) to guarantee sufficiently high SNR ($SNR > T$). At the same time, RNs need to be allocated conservatively, in order to reducing transmission conflicts, as already discussed in subsection \ref{refsce}. Section \ref{relaynode} provides an   optimum algorithm to this end. In the second step, and considering all candidate paths, we focus on interference avoidance. To this end, we provide a heuristic to define  a transmission scheduling diagram which can eliminate interference (Section \ref{schedulingtrans}). The third and final step maximizes the network throughput by choosing the optimal path for each traffic demand, considering the already determined set of candidate paths with interference avoidance. The related optimization method is presented in Section \ref{MILP}. 

\subsection{Finding candidate paths}\label{relaynode}
\par Finding candidate paths is concerned with finding paths that satisfy $SNR > T$ (threshold value) between one source node (BS node $b \in B$) and one destination node (UE node $u \in U$). Let us define $P_{\mathbb{G}^{'}}$ as a set of all candidate paths without relay nodes between $b$ and $u$, which include all paths $P_a \in P_{\mathbb{G}^{'}}$ with $SNR > T$ added to the set $\mathbb{P}_{\gamma}$. For all other paths,  $P_i \in P_{\mathbb{G}^{'}}$ with $SNR \leq T$, we consider including relay nodes $e$, such that the resulting $SNR > T$. Also these paths are included in the set of all candidate paths, i.e., $\mathbb{P}_{\gamma}$.  Algorithm \ref{algorithm2} minimize the number of relay nodes $e$ used in $P_i$. Reducing the number of used relay nodes is critical not only to efficient resource allocation, but also because they are known to contributing to the co-channel interference. 

\par The input of $Finding\_RelayNode$ function in Algorithm \ref{algorithm2} includes SNR threshold $T$ and path $P_i \in P_{\mathbb{G}^{'}}$. Let $be$ be the transmitting node variable considered, where $be$ can be BS or relay node. At the beginning, since there is no relay node $e$ found, variable $be$ is assigned by the BS of path $P_i$ (line \ref{a1}). We consider all hops of path $P_i(be,UE)$ between $be \gets P_i(BS)$ and UE (starting from the last hop) (line \ref{a5}), and choose one relay node belonging to the considered hop which is closest to UE (line \ref{a6}). If a relay node with $SNR_{(be,rn)}> T$ is found (line \ref{a7}), then $rn$ becomes the new $be$ (line \ref{a9}), and it is also added to path $P_i$ at the hop under consideration (line \ref{a10}). By selecting relay nodes closest to UE, we can minimize the number of relay nodes used for each path. We then check if $SNR_{(be,P_i(UE))}>T$ between the newly added relay node ($be \gets rn$) and UE of path $P_i$ (line \ref{a11}). If yes, we  found a candidate solution and stop the search (lines \ref{a12}-\ref{a13}). Otherwise, we restart the iteration, now searching for a new relay node in the path between the last added relay node and UE (line \ref{a5}). In case that viable relay nodes cannot be found,  $P_i \gets \emptyset$ and we stop the search (lines \ref{a18}-\ref{a19}). 

The output of Algorithm \ref{algorithm2} can be either the new $P_i$ including one or more relay nodes, which will then be added to $\mathbb{P}_{\gamma}$, or an empty path set.

\begin{algorithm}
\caption{Finding\_RelayNode($T$, $P_i$)}
\label{algorithm2}
\begin{algorithmic}[1]
\State $be \gets P_i(BS)$;\label{a1}
\State $iterate \gets True$;\label{a2}
\While {iterate} \label{a3}
  \State $viable \gets False$; \label{a4}
  \For {each $hop$ from $P_i(be,UE)$, starting from last} \label{a5}
    \State $rn \gets$ closestRelayNode($hop$);\label{a6}
    \If {$SNR_{(be,rn)} > T$}\label{a7}
      \State $viable \gets True$;\label{a8}
      \State $be \gets rn$;\label{a9}
      \State $P_i \gets rn$;\label{a10}
      \If {$SNR_{(be, P_i(UE))} > T$}\label{a11}
        \State $iterate \gets False$;\label{a12}
         \State break;\label{a13}
      \EndIf \label{a14}
     
    \EndIf\label{a15}
  \EndFor \label{a16}
  \If {$not$ $viable$} \label{a17}
    \State $P_i \gets \emptyset$; \label{a18}
    \State break;  \label{a19}
  \EndIf \label{a20}
\EndWhile \label{a21}
\State \textbf{OUTPUT}: $P_i$; \label{a22}
\end{algorithmic}
\end{algorithm}

\subsection{Transmission scheduling to avoid conflicts} \label{schedulingtrans}
To eliminate interference among paths in $\mathbb{P}_{\gamma}$, we propose a simple transmission scheduling heuristic. Let us consider an illustrative example of transmission scheduling. Consider a path in Fig. \ref{genscenario}  BS from $0$ $\rightarrow$ RIS $1$ $\rightarrow$ RIS $2$ $\rightarrow$ UE $0$. We note that three other paths can cause interference on that path, including: (i) The cylindrical beam of path BS $2$ $\rightarrow$ RIS $3$ $\rightarrow$ RIS $1$ $\rightarrow$ RIS $4$ $\rightarrow$ UE $2$ causing interference on RIS $1$; (ii) the conical beam of path BS $1$ $\rightarrow$ UE $1$ causing interference on RIS $2$; and (iii) the conical beam of BS $3$ $\rightarrow$ UE $3$ causing interference on UE $0$. Each interference value can be calculated from Eq. \eqref{interequ}, and so we have three corresponding values, $\mathbb{I}_{(be^\prime,eu)}^1$, $\mathbb{I}_{(be^\prime,eu)}^2$, and $\mathbb{I}_{(be^\prime,eu)}^3$. Let us assume that $\mathbb{I}_{(be^\prime,eu)}^1 \leq \mathbb{I}_{(be^\prime,eu)}^2 \leq \mathbb{I}_{(be^\prime,eu)}^3$. If $\mathbb{I}_{(be^\prime,eu)}^1$ is applied to Eq. \eqref{SNIRvi}, we can verify if $SNIR_{(be,eu)} < T$, if so, we know that all four paths cannot be simultaneously scheduled. If $SNIR_{(be,eu)} > T$, on the other hand, we can continue adding the interference of the next interfering transmission from the ordered set until we find the maximal subset of paths that can be scheduled along with the first scheduled path with $SNIR_{(be,eu)} \geq T$. This we denote as the transmission schedule diagram.
%path BS $2$ $\rightarrow$ RIS $3$ $\rightarrow$ RIS $1$ $\rightarrow$ RIS $4$ $\rightarrow$ UE $2$ and the interfered transmission can be simultaneously scheduled due to significant interference value. Then, we keep calculating $\mathbb{I}_{total}=\mathbb{I}_{total}+\mathbb{I}_{(be^\prime,eu)}^2$. Now, if $SNIR_{(be,eu)} = \frac{P_{eu} \cdot G_{be} \cdot G_{eu}}{k \cdot T \cdot W +\mathbb{I}_{total}} \leq T$, the two remaining interfering transmissions and the interfered transmission cannot be simultaneously scheduled, and we stop calculating transmission scheduling. Otherwise, the interfering transmission BS $1$ $\rightarrow$ UE $1$ and the interfered transmission can be be simultaneously scheduled. Similarly, we consider the third interfering transmission with $\mathbb{I}_{total}=\mathbb{I}_{total}+\mathbb{I}_{(be^\prime,eu)}^3$, and end the processing of scheduling transmission for the interfered transmission.

Let us assume that each path transmits between $(be,\{I_{(be,eu)}\cup eu\})$ the amount of traffic $y_{(be,eu)}$. We define the time for transmitting $y_{(be,eu)}$ as $\frac{y_{(be,eu)}}{C_{(be,eu)}}$, whereby $C_{(be,eu)}$ is the channel capacity between node $be$ and node $eu$ of each transmission, i.e., 
\begin{equation}\label{Cbeeu}
C_{(be,eu)}=W \cdot log_2(1+SNR_{(be,eu)}).
\end{equation}

%By scheduling transmissions to eliminate interference, any paths and their related traffic transmissions that cannot be simultaneously scheduled are conflicted, thus at most one of them can be active at any given time.

%\subsection{Algorithm of finding end-to-end paths}\label{algo12}

%\subsection{Mixed Integer Linear Programming (MILP)}\label{lpm}
\begin{table}[t!]
  \centering
    %\vspace{0.2 cm}
  \caption{List of input parameter and variable notations.}
  \label{tab:table2}
  \begin{tabular}{ll}
    \toprule
 Input param. & Meaning\\
    \midrule
    %{\footnotesize
    $\theta$ & Application traffic set. \\
    $k$ & Each traffic $\gamma \in \theta$ has $k$ Gbits\\
    $\mathbb{P}_{\gamma}$ & Set $\mathbb{P}_{\gamma}$ of considered paths to send traffic $\gamma$  from one \\ &  BS to one UE.\\
   $C_{(be,eu)}$ & Channel capacity between node $be$ and node $eu$.\\
   %}
\toprule
  Variables & Meaning\\
  \midrule
  %{\footnotesize
 $f^P_{\gamma}$ & Binary routing variable belonging to traffic $\gamma$ chosen to\\ &   send over path $P \in \mathbb{P}_{\gamma}$. \\
 $\lambda$ & Throughput Controlling Coefficient (TCC).\\
 $y_{(b,eu)}$  & Traffic variable for transmission between BS $b$ and relay \\ & node $e$ or UE node $u$, i.e., node $eu$.\\
  $y_{(e,eu)}$  & Traffic variable for transmission between relay node $e$   \\ & and another relay  $e$ or UE $u$, i.e., network node $eu$.\\
%}
\bottomrule
  \end{tabular}%\vspace{-0.5cm}
\end{table}

\subsection{Throughput maximization}\label{MILP}
To maximize throughput, we use the set of candidate paths $P_{\gamma}$ from Section \ref{relaynode} and the transmission scheduling diagram from Section \ref{schedulingtrans} as input for a MILP model. In the MILP model, traffic of $k$ symbols is sent over the best single path (i.e., path with least conflicts) selected from  $P_{\gamma}$, with the goal of maximizing the connection throughput. All input parameters for the model are summarized in Table \ref{tab:table2}. 

Let us again explain the basic idea with an example. Assume that BS $0$ wants to send its data to UE $0$ over the best single path chosen from a set of two possibly shortest paths: The first path BS $0$ $\rightarrow$ RIS $6$ $\rightarrow$ RN $0$ $\rightarrow$ RIS $7$ $\rightarrow$ UE $0$ in Fig. \ref{genscenario} may only conflict with one transmission BS $3$ $\rightarrow$ UE $3$, while the second path BS $0$ $\rightarrow$ RIS $1$ $\rightarrow$ RIS $2$ $\rightarrow$ UE $0$ may be conflicted with up to three other paths. Therefore, the first path in the example will be selected due to less conflicts, although the transmission distance of the first path (in terms of hop count) is longer, thus maximizing the amount of time for scheduled transmissions, i.e., maximizing the network throughput. We will later refer to this path optimization strategy as \emph{least interference path} and contrast it to the method  where traffic is routed over the shortest path only (the second path in the example), i.e.,  $|P_{\gamma}|=1$. As per our analyses, the shortest path is likely to exhibit more conflicts,  thus reducing the connection throughput.

Let us define Throughput Controlling Coefficient (TCC) as a maximum multiplier $\lambda$ such that when each traffic demand of $k$ Gbits is multiplied by $\lambda$,  it can be transmitted with least interference. If $\lambda > 1$, it can accommodate more connections $\gamma$ with more traffic; i.e., only part of the available resources is being used. If $\lambda = 1$, the network resources are used optimally, i.e., without wasting resources. Otherwise, for $\lambda < 1$, the network cannot accommodate the traffic requests $\gamma$. The traffic allocation that maximizes the throughput by minimizing interference can be formulated as a MILP model. The output of MILP model generates the best THz downlink path selection for each traffic demand $\gamma$ of $k$ Gbits by calculating $f_{\gamma}^P$ binary variables and variable $\lambda$. The objective function is given as:
\begin{equation} \nonumber
\text{Maximize} \; \lambda 
\end{equation}
\begin{equation}
\begin{split}
\text{s.t. } {\displaystyle \sum\limits_{P \in \mathbb{P}_{\gamma}}f_{\gamma}^P = 1, \; \forall \; \gamma \in \theta}. \label{cont1}
\end{split}
\end{equation}
\begin{equation}
\begin{split}
y_{(b,eu)} =  \sum_{\gamma \in \theta} \; \sum_{P \in \mathbb{P}_{\gamma};t_{eu}^b \in P}  f_{\gamma}^P \cdot k, \; \forall \; b \in B, eu \in \{E \cup U \}. \label{cont2}
\end{split}
\end{equation}
\begin{equation}
\begin{split}
 y_{(e,eu)} = \sum_{t^{be}_e \in T_{e}^{in}}  \; \sum_{\gamma \in \theta}   \; & \sum_{P \in \mathbb{P}_{\gamma};t^{be}_e t^e_{eu} \in P} f_{\gamma}^P \cdot k, \; \\ & \forall \; e \in E, eu \in \{E \cup U \}. \label{cont3}
\end{split}
\end{equation}
%\begin{equation}
%\begin{split}
%\sum_{t^{be}_e \in T_{e}^{in}} \; \sum_{\tilde{a} \in \tilde{A}} \; \sum_{g_{\tilde{a}} \in G_{\tilde{a}}} \; & \left[ \sum_{o^{\tilde{a}}_{g_{\tilde{a}}} \in \mathbb{O}^{\tilde{a}}_{g_{\tilde{a}}}} \; \sum_{P \in \mathbb{P}_{o^{\tilde{a}}_{g_{\tilde{a}}}}} f_{o^{\tilde{a}}_{g_{\tilde{a}}}}^P+ \right. \\ & \left. \sum_{r^{\tilde{a}}_{g_{\tilde{a}}} \in \mathbb{R}^{\tilde{a}}_{g_{\tilde{a}}}} \; \sum_{P \in \mathbb{P}_{r^{\tilde{a}}_{g_{\tilde{a}}}}} f_{r^{\tilde{a}}_{g_{\tilde{a}}}}^P \right] \leq L, \; \forall \; e \in E.  \label{cont4}
%\end{split}
%\end{equation}
\begin{equation}
\begin{split}
\sum_{(be,\{I_{(be,eu)}\cup eu\}) \in C_{\mathbb{F}}} \frac{y_{(be,eu)}}{C_{(be,eu)}} \leq \frac{1}{\lambda}, \; \forall \; C_{\mathbb{F}} . \label{cont5}
\end{split}
\end{equation}

Constraint \eqref{cont1} defines that for each traffic demand $\gamma \in \theta$ only one single path $P$ can be chosen from the set $\mathbb{P}_{\gamma}$ of paths between BS source node and one UE destination node. Constraint \eqref{cont2} guarantees that the total amount of traffic $y_{(b,eu)}$ between BS node $b\in B$ and its destination node $eu \in \{E\cup U\}$ is sent over path $(b,\{R_{(be,eu)}\cup eu\})$ (transmission $t^b_{eu}$ for short) from all traffic demands $\gamma \in \theta$.  Constraint \eqref{cont3} guarantees that for each transmission $(e,\{R_{(be,eu)}\cup eu\})$ (with transmission $t^e_{eu}$ for short), its total amount of traffic $y_{(e,eu)}$ between the prospective relay node $e$ and the related destination node is coming from all traffic demands $ \theta$ from the relay node $e$ on the path $(e,\{R_{(be,eu)}\cup eu\})$.  %The set of previous transmissions $T_e^{in}$ go in its considered relay node $e$.

%\hl{Constraint} \eqref{cont3} guarantees that the total amount of traffic $y_{(e,eu)}$ between each relay node $e$ and its destination node $eu \in \{E\cup U\}$ is sent over path $(e,\{R_{(be,eu)}\cup eu\})$ (with transmission $t^e_{eu}$ for short) for all traffic demands $\gamma \in \theta$ whose previous transmission $t^{be}_e$ between its previous transmitting node $be$ and it, where $t^{be}_e \in T^{in}_e$ (set of incoming transmissions  at relay node $e$).

To explain constraint \eqref{cont5} related to transmission scheduling, let us use an example from Fig. \ref{genscenario}. Since paths BS $0$ $\rightarrow$ RIS $1$ $\rightarrow$ RIS $2$ $\rightarrow$ UE $0$ with traffic $y_{(be_0,eu_0)}$  and BS $1$ $\rightarrow$ UE $1$ with $y_{(be_1,eu_1)}$  are conflicted, at most one of them can be active at any given time.
%, i.e., consider in one second, $\frac{y_{(be_0,eu_0)}}{C_{(be_0,eu_0)}}+\frac{y_{(be_1,eu_1)}}{C_{(be_1,eu_1)}} \leq 1$.
Hence, we consider $C_{\mathbb{F}}$ to be the set of conflicted paths in the entire network, or as we called in Section \ref{schedulingtrans}, the transmission schedule diagram. Therefore, in Fig. \ref{genscenario}, we have $11$ different sets $C_{\mathbb{F}}$: $\{$ BS $2$ $\rightarrow$ RIS $3$ $\rightarrow$ RIS $1$ $\rightarrow$ RIS $4$ $\rightarrow$ UE $2$; BS $0$ $\rightarrow$ RIS $1$ $\rightarrow$ RIS $2$ $\rightarrow$ UE $0$ $\}$, $\{$ BS $0$ $\rightarrow$ RIS $1$ $\rightarrow$ RIS $2$ $\rightarrow$ UE $0$; BS $4$ $\rightarrow$ UE $4$ $\}$, $\{$ BS $0$ $\rightarrow$ RIS $1$ $\rightarrow$ RIS $2$ $\rightarrow$ UE $0$; BS $3$ $\rightarrow$ UE $3$ $\}$, $\{$ BS $0$ $\rightarrow$ RIS $1$ $\rightarrow$ RIS $2$ $\rightarrow$ UE $0$; BS $1$ $\rightarrow$ UE $1$ $\}$, $\{$ BS $3$ $\rightarrow$ UE $3$; RN $0$ $\rightarrow$ RIS $7$ $\rightarrow$ UE $0$ $\}$, $\{$ BS $7$ $\rightarrow$ UE $7$; BS $5$ $\rightarrow$ RIS $5$ $\rightarrow$ RN $5$ $\}$, $\{$ BS $8$ $\rightarrow$ RIS $3$ $\rightarrow$ UE $8$; BS $2$ $\rightarrow$ RIS $3$ $\rightarrow$ RIS $1$ $\rightarrow$ RIS $4$ $\rightarrow$ UE $2$ $\}$, $\{$ BS $0$ $\rightarrow$ RIS $6$ $\rightarrow$ RN $0$ $\}$, $\{$ BS $6$ $\rightarrow$ RN $6$ $\}$, $\{$ RN $6$ $\rightarrow$ UE $6$ $\}$, and $\{$ RN $5$ $\rightarrow$ UE $5$ $\}$. Since there are conflicting paths in every set $C_{\mathbb{F}}$, at most one transmission on a path $(be,\{R_{(be,eu)}\cup eu\})$ can be active at a given time, and the fraction of active time of that transmission with traffic value $y_{(be,eu)}$ is $\frac{y_{(be,eu)}}{C_{(be,eu)}}$, where $C_{(be,eu)}$ is the channel capacity per Eq. \eqref{Cbeeu}. Based on  \cite{viennew}, we can now infer that the constraint for scheduling these transmissions can be expressed as:
\begin{equation}\label{constraintsche}
\sum_{(be,\{I_{(be,eu)}\cup eu\}) \in C_{\mathbb{F}}} \frac{y_{(be,eu)}}{C_{(be,eu)}} \leq 1, \; \forall \; C_{\mathbb{F}},
\end{equation}  
It should be noted that as we scale each traffic $\gamma$ by $\lambda$, we get the constraint \eqref{cont5}. As a result, the MILP model uses this information from the sets of $C_{\mathbb{F}}$ with the constraint \eqref{cont5} to find the best single path with least conflicts.

%%%%%%%%%%%%%%%%%%%%%%%% Performance evaluatino
%%%%%%%%%%%%%%%%%%%%%%%% Performance evaluatino
\section{Performance evaluation}\label{perev}
\begin{figure*}[ht]
\captionsetup[subfigure]{labelformat=empty}
  \centering
\subfloat[]{\includegraphics[ width=7cm, height=5cm]{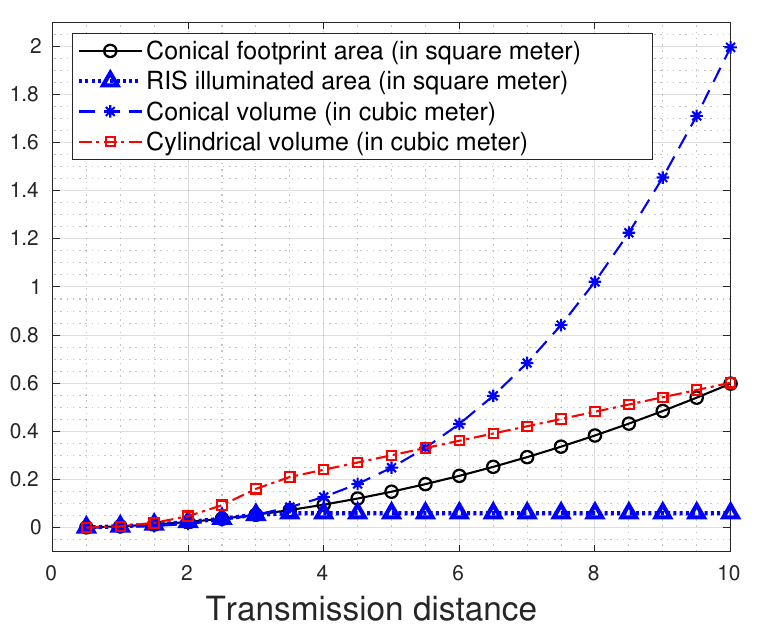}
  \label{beam_coverage_5_degree}}
  \subfloat[]{\includegraphics[ width=7cm, height=5cm]{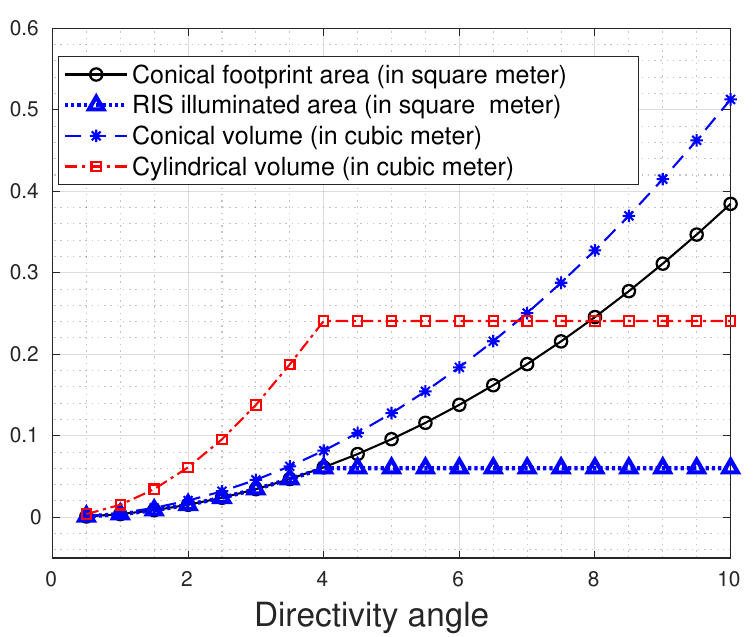}
  \label{beam_coverage_4_meters}}
  \caption{Analysis of beam shapes, where (a): left figure, (b): right figure.}
  \label{beamshaperesult}
  %\vspace{-0.67cm}
  \end{figure*}
  
\par In this section, we first verify the beam shape and interference model, and then study impact of interference effect on network performance. We also show results that demonstrate that throughput can be maximized by interference avoidance algorithm in relay mesh THz network. 
  
\par Fig. \ref{beamshaperesult} shows the results related to the beam shapes generated from one BS or one RIS. The dimensions $x$ and $y$ of RIS reflecting elements are $dx=dy=0.0024$ m, and the number of reflecting elements $N=10453$ elements (parameters taken from \cite{9386246}). Let us first assume that BS sends the signal to UE via one RIS. For simplicity, we assume that the threshold distance is $d_{th}=d_1+d_2=2*d_1$, where $d_1$ is the transmission distance of the first hop (between BS and RIS), and $d_2$ is the maximum of transmission distance from RIS to UE. In Fig. \ref{beam_coverage_5_degree}, we consider the directivity angle of $5$ degrees and transmission distance $d_1$ and $d_2$ from $0.5$ to $10$ m in step of $0.5$. In Fig. \ref{beam_coverage_4_meters}, we consider transmission distance of $d_1=d_2=4$ m, and the directivity angle from $0.5$ degree to $10$ degrees in steps of $0.5$. Eqs. \eqref{radius}, \eqref{Sira}, \eqref{Sfinal}, \eqref{Ractualill}, \eqref{cone1}, \eqref{cylinder1}, \eqref{cylinder2} in section \ref{beamshape} are used to calculate the values shown results in Fig. \ref{beamshaperesult}. We can observe that with increasing transmission distance and directivity angle, the conical footprint area as well as conical volume increases. We also observe  that although the conical foot area increases with increasing the transmission distance and directivity angle, RISs cannot capture the entire beam. Therefore, the results for RIS illuminated area (Fig. \ref{beam_coverage_4_meters}), cylindrical volume (Fig. \ref{beam_coverage_4_meters}), or RIS illuminated area (Fig. \ref{beam_coverage_5_degree}) remain constant with larger directivity angles. However, the cylindrical volume (Fig. \ref{beam_coverage_5_degree}) increases slowly because the transmission distance increases. The analysis of beam coverage is very critical, as it can help us  identify the corresponding interference areas.

\begin{figure*}[ht]
\captionsetup[subfigure]{labelformat=empty}
  \centering
  \subfloat[]{\includegraphics[ width=7cm, height=5cm]{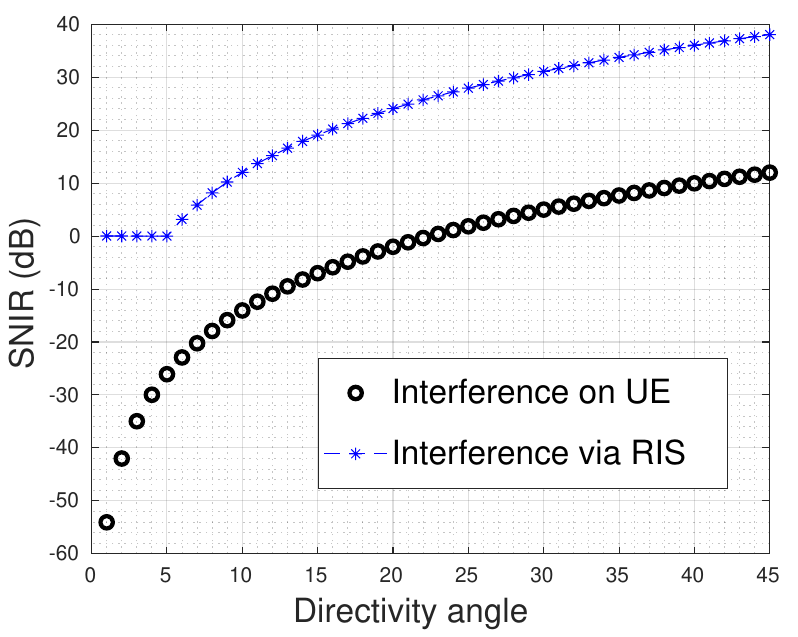}
  \label{interre}}
  \subfloat[]{\includegraphics[ width=7cm, height=5cm]{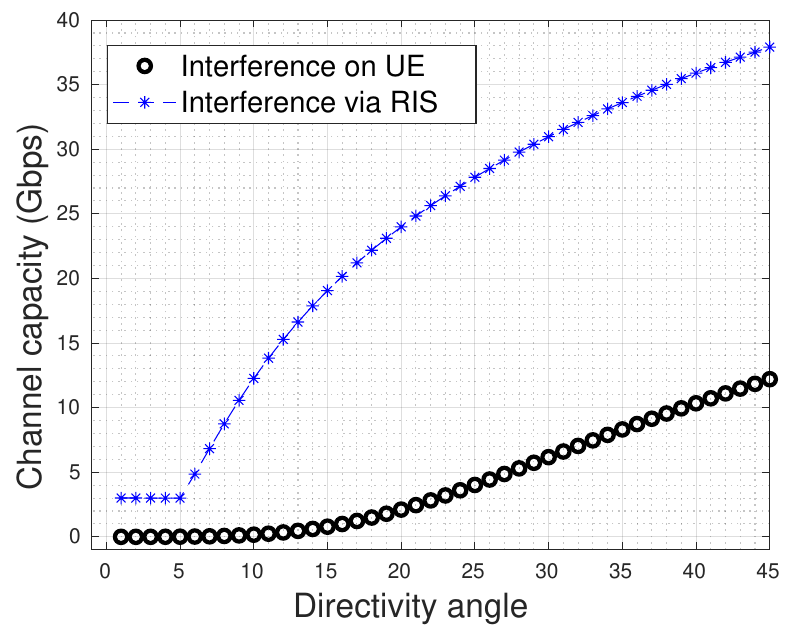}
  \label{capacitych}}
  \caption{Interference analysis and channel capacity, where (a): left figure, (b): right figure}
  \label{intercap}
  %\vspace{-0.67cm}
  \end{figure*}

  \begin{figure}[ht]
%\vspace{0.1 cm}
\centering
{\includegraphics[ width=6.4cm, height=4.4cm]{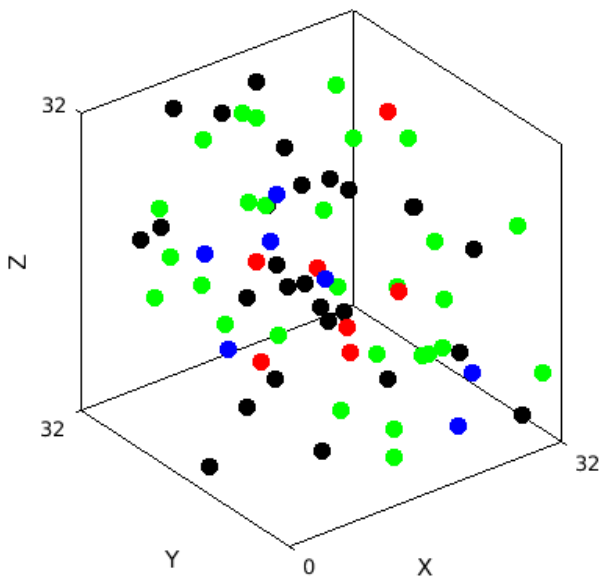}}
\caption{THz relay mesh network as 3D area $(32,32,32)$. } \label{topology}
%\vspace{-0.68 cm}
\end{figure}

In Fig. \ref{intercap}, we analyze the impact of beam shape of the paths that cause interference on the other existing paths. We numerically analyze examples earlier illustrated in  Fig. \ref{genscenario}. Path BS $8$ $\rightarrow$ RIS $3$ $\rightarrow$ UE $8$ suffers interference by path with conical beam BS $2$ $\rightarrow$ RIS $3$ $\rightarrow$ RIS $1$ $\rightarrow$ RIS $4$ $\rightarrow$ UE $2$ on RIS $3$, as does path BS $5$ $\rightarrow$ RIS $5$ $\rightarrow$ RN $5$ by BS $7$ $\rightarrow$ UE $7$ on RN $5$. Let us assume the following parameters to analyze: THz frequency $1$ THz, tranmssion power $10$ watt, bandwidth $3$ GHz; the values of $d_x$, $d_y$, and number of reflecting elements are taken from the previous example. The distances BS $2$ $\rightarrow$ RIS $3$, BS $8$ $\rightarrow$ RIS $3$, RIS $3$ $\rightarrow$ UE $8$, BS $5$ $\rightarrow$ RIS $5$, and RIS $5$ $\rightarrow$ RN $5$ are $1$ m. The distance BS $7$ $\rightarrow$ RN $5$ is $2$ m. Assume that RN $5$ captures the whole interfering beam from BS $7$. The directivity angle of interfered transmission from BS $5$ and BS $8$ is fixed to $5$ degrees. We apply the analysis of SNIR with Eq. \eqref{SNIRvi} and Eq. \eqref{interequ}, Section \ref{beamshape}. Here, x-axis is the directivity angle of interfering transmission from BS $2$ and BS $7$ in ranges  from $1$ degree to $45$ degrees in steps of $1$. In Fig. \ref {interre}, SNIR of interference via RIS is clearly better due to the fact that RIS does not capture the entire footprint of the interfering conical beams when the directivity angle increases. On the other hand, we see that with small directivity angle $\leq 5$ degrees, RIS SNIR remains constant due to the rather small number of reflecting elements captured. This is because when we increase the antenna gain (i.e., decreasing directivity angle),  the number of reflecting elements $|N^\prime|$ captured by RIS decrease (refer back to the analysis from Eqs. \eqref{radius}, \eqref{Sira}, \eqref{Sfinal}, \eqref{Ractualill}, \eqref{Ncomma} in Section \ref{beamshape}). Based on Eq. \eqref{eqex1}, Eq. \eqref{eqex2} (in Section \ref{SNRend}), and Eq. \eqref{SNIRvi}, Eq. \eqref{interequ} (in Section \ref{beamshape}), we can see that interference decreases with decreasing  number of interfering reflecting elements on RIS. The interfering cylindrical beams exhibit the same behaviors, not shown here for simplicity. Fig. \ref{capacitych} shows the impact of interference on the channel capacity applied from Eq.\eqref{Cbeeu} in Section \ref{schedulingtrans}. Clearly, the performance of RIS is better in terms of interference reduction.

%\hl{this case study is not clear at all. i do not know what is studied. if this is related to analyses. both figures and all three curves need to be refered abck o the section and the exact equation. Where are these 6 equations? }

\begin{figure}[ht]
%\vspace{0.1 cm}
\centering
{\includegraphics[ width=7.4cm, height=5cm]{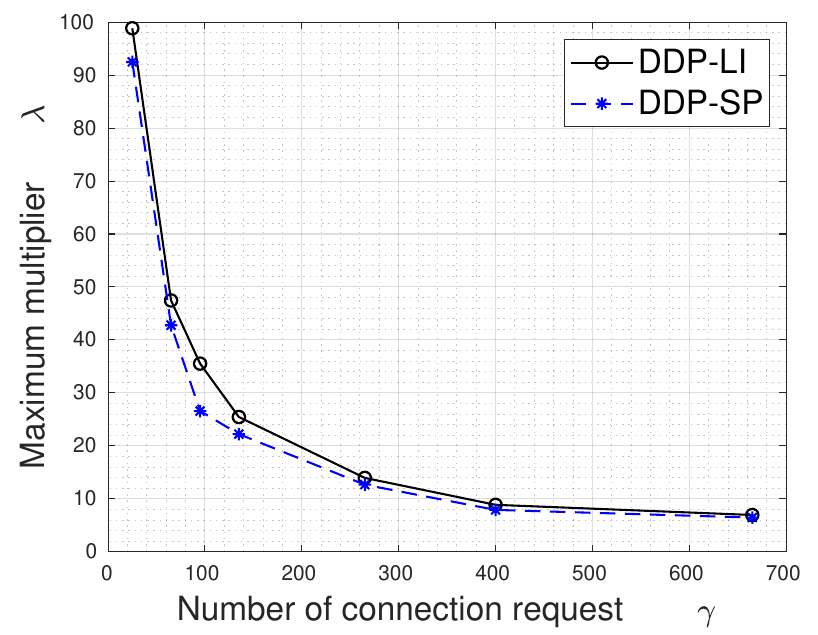}}
\caption{Maximum multiplier vs. connection requests } \label{lamda}
%\vspace{-0.68 cm}
\end{figure}

\begin{figure}[ht]
%\vspace{0.1 cm}
\centering
{\includegraphics[ width=7.4cm, height=5cm]{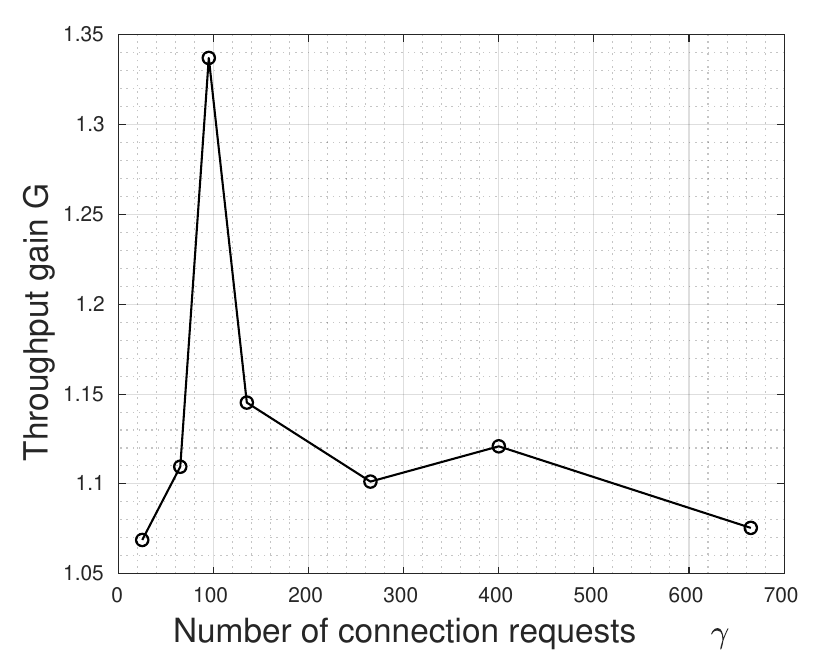}}
\caption{Throughput gain.} \label{throughputgain}
%\vspace{-0.68 cm}
\end{figure}

The last set of results focuses on a network generated as a 3D area of size $32 \times 32 \times 32$ m (Fig. \ref{topology}). The simulation parameters are summarized in Table \ref{tab:table1}.  We assume that BS source nodes send their data over path reach of $20$ m at most. For longer paths, we deploy RISs and relay nodes  towards the destination nodes UE, as proposed earlier. We generate a traffic matrix for each pair of  transmitting nodes (BSs) and receiving nodes (UEs), with the uniform traffic request $\gamma$. 

Fig. \ref{lamda} analyzes the $\lambda$ values for the TCC variable . We compare the two path selection methods: least interference (best) path selected from the set of candidate paths (we refer to it as Downlink Distribution Path with Least Interference, DDP-LI), and the shortest path out of all candidate paths (DDP-SP). With DDP-LI, we send the traffic request over the best path chosen from the set $\mathbb{P}_{\gamma}$ of $5$ candidate paths. DDP-SP always sends traffic over the shortest path from the set of candidate paths albeit with possibly larger interference effect, i.e., $|\mathbb{P}_{\gamma}|=1$. The traffic requested per connection is varied as  $\gamma$ (x-axis): $25,65,95,135,265,400,665$. Based on the MILP algorithm from Section \ref{MILP}, we now show the results obtained on the $y$-axis as the output values $\lambda$.

%\hl{Where are these two methods explained i??? calling it sisngle path is meaningless since they are always single so this is a bad name. but SDDT-Sh is a problem since we nowhere choose shortest path in seciton IV< so what is the idea here????? also these results need to focus on maximinzing THROUHPut while minimizing interefence? where is this shown???????}

 \begin{table}[t!]
  \centering
%\vspace{0.4 cm}
  \caption{Parameters used in simulations.}
  \label{tab:table1}
  \begin{tabular}{ll}
    \toprule
   Values & Meaning\\
    \midrule
    $k=0.05$ & Traffic $\gamma \in \theta$ in Gbit.\\
    $|B|=7$ & Number of BS source nodes.\\
    $|U|=7$ & Number of UE destination nodes.\\
    $|R|=28$ & Number of intermediate RIS nodes.\\
    $|E|=28$ & Number of relay nodes.\\
    $\alpha=15$ & antenna directivity angle in degree.\\
    $k(f)=0.0016$ & Molecular absorption coefficient in $m^{-1}$.\\
    $W=3$ & Bandwidth in GHz.\\
    $f=1$ & Operational center frequency.\\
    $P_{be}=1$ & Power of BS or relay node in Watt.\\
    $T_o=300$ & Temperature in Kelvin.\\
    $T=10$ & Threshold of SNR in dB.\\
    $d_x,d_y=\frac{c}{2f}$ & $x$ and $y$ dimensions of each reflecting element.\\
    $S_{RIS}=0.0022$ & RIS area in $m^2$.\\
    $|\mathbb{P}_{\gamma}|=5$ & Set of candidate shortest paths considered for DDP-LI.\\
    $|\mathbb{P}_{\gamma}|=1$ & The shortest path used in DDP-SP.\\
     \bottomrule
  \end{tabular}%\vspace{-0.76cm}
\end{table}

As expected, Fig. \ref{lamda} confirms that DDP-LI achieves a better end-to-end network throughput (value $\lambda$) than DDP-SP.  The throughput decreases with an increasing traffic $\gamma$ because it is more difficult to find paths without conflicts.  With the value $\lambda$, we have the maximum traffic achievable of each traffic demand $\gamma$ sent over the THz system, whereby $k_{max}= k \cdot \lambda$. Therefore, if $\lambda \geq 1$, for each transmission with $k$ Gbit, a valid path can be successfully selected in the network.  

Although in Fig. \ref{lamda} we for clarity show a maximum of 665 connection requests $\gamma$, we did some evaluations with $4665$, $4935$, and $5065$ traffic demands $\gamma$. With $4665$ and $4935$ traffic $\gamma$, DDP-LI still satisfies the throughput value $\lambda > 1$, while DDP-SP is down to $\lambda < 1 $. With $5065$ traffic $\gamma$, all methods yield $\lambda < 1 $. In scenario where adequate paths cannot be found, the algorithm can hence decide if the traffic demand is blocked due to insufficient QoS (not studied here).

Fig. \ref{throughputgain} shows the throughput gain $G$ defined as the ratio of the TCC value obtained by the DDP-LI in relation to the TCC value obtained by the DDP-LI, i.e., $G = \frac{\lambda_{DDP-LI}}{\lambda_{DDP-SP}}$. We calculate the throughput gain $G$ in Fig. \eqref{throughputgain} from the values in Fig. \ref{lamda}. With DDP-LI,  the throughput gain is significant. When the connection requests increase up to $95$, the DDP-LI exhibits many options to avoid conflicts, while DDP-SP is limited by one shortest path only. Therefore, the throughput of DDP-LI  peaks. The throughput gain of DDP-LI decreases significantly with the increasing connection request, due to the fact that in more loaded networks it is harder to find   paths without conflicts. However,  DDP-LI remains a valid choice for reducing conflicts and maximizing the network throughput in low to moderately loaded networks.
%\hl{how do you explain the spike in thoughput gain at value 100?, make these figures one after another in a column so they can be bigger and leggible.}
%\hl{this text above is full of garbage left over from previous version. please fix and focus on trhoughput maximinzation and itnerefence avoidance iwth MILP here and NOT on old resutls that have nothing to do ith seciton IV. this needs to show how MILP maximinzes the throughput and minimizies interefence. so this should be seen from the resutls. }

%%%%%%%%%
\section{Conclusion}\label{conl}
In this study, we addressed the crucial matter of interference avoidance in Terahertz (THz) relay mesh networks, introducing methods aimed at maximizing network throughput. Our first step was to conduct an innovative analysis of geometric beam steering shapes, specifically conical and cylindrical, within various transmission scenarios. We then developed routing and path selection strategies that effectively avoid interference. Alongside this, we designed and assessed an efficient scheduling transmission algorithm suited to the network system, optimizing both throughput and path selections, all while minimizing the allocation of relay nodes. Our findings clearly indicated that our approach to routing interference avoidance could significantly maximize throughput in comparison to traditional shortest path methods. We further demonstrated that as transmission distance and directivity angle increased, so too did the conical footprint area and conical volume. However, it was observed that, despite the increase in the conical foot area with the elevation of transmission distance and directivity angle, Retroreflective Interference Surfaces (RISs) are unable to capture the entirety of the beams. Our findings provide critical insights into the nature and extent of interference coverage, and more importantly, how it can be avoided. This research opens pathways for further work in this field, underscoring the importance of interference avoidance in THz relay mesh networks to achieve optimal network performance.

\bibliographystyle{IEEEtran}

\bibliography{nc-rest}

\end{document}